\documentclass[aps,prb,twocolumn,amsmath,amssymb,floatfix,superscriptaddress,reprint]{revtex4-1}

\usepackage{graphicx}
\usepackage{dcolumn}
\usepackage{bm}
\usepackage[bookmarks,bookmarksopen,bookmarksnumbered,colorlinks,linkcolor=red,linktocpage,citecolor=blue,urlcolor=cyan,pdfpagemode=UseOutline]{hyperref}

\usepackage[utf8]{inputenc}
\usepackage[T1]{fontenc}
\usepackage{mathptmx}
\usepackage{etoolbox}

\def\ben{\begin{equation}}
\def\een{\end{equation}}

\DeclareMathOperator{\erf}{erf}
\newcommand{\parref}[1]{(\ref{#1})}

\newcommand{\IB}{I_\mathrm{B}}

\newcommand{\revise}[1]{{#1}}

\begin{document}

\title{The mechanism of the \revise{irradiation synergistic effect of Silicon bipolar junction transistors} explained by multiscale simulations of Monte Carlo and excited-state first-principle calculations}
\author{Zeng-hui Yang\revise{$^*$}}
\email{yangzenghui_mtrc@caep.cn}
\affiliation{Microsystem and Terahertz research center, China Academy of Engineering Physics, Chengdu, China 610200}
\affiliation{Institute of Electronic Engineering, China Academy of Engineering Physics, Mianyang, China 621000}
\author{Yang Liu}
\affiliation{Microsystem and Terahertz research center, China Academy of Engineering Physics, Chengdu, China 610200}
\affiliation{Institute of Electronic Engineering, China Academy of Engineering Physics, Mianyang, China 621000}
\author{Ning An}
\affiliation{Microsystem and Terahertz research center, China Academy of Engineering Physics, Chengdu, China 610200}
\affiliation{Institute of Electronic Engineering, China Academy of Engineering Physics, Mianyang, China 621000}
\author{Xingyu Chen}
\affiliation{Microsystem and Terahertz research center, China Academy of Engineering Physics, Chengdu, China 610200}
\affiliation{Institute of Electronic Engineering, China Academy of Engineering Physics, Mianyang, China 621000}
\date{\today}

\begin{abstract}
Neutron and $\gamma$-ray irradiation damages to transistors are found to be non-additive, and this is denoted as the irradiation synergistic effect (ISE). Its mechanism is not well-understood. The recent defect-based model [ACS Appl. Electron. Mater. 2, 3783 (2020)] for Silicon bipolar junction transistors (BJT) achieve quantitative agreement with experiments, but \revise{its assumptions on the defect reactions are unverified}. \revise{Going beyond the model} requires directly representing the effect of $\gamma$-ray irradiation in first-principles calculations, which is not feasible previously. In this work, we examine the defect-based model of the ISE by developing a multiscale method for the simulation of the $\gamma$-ray irradiation, where the $\gamma$-ray-induced electronic excitations are treated explicitly in excited-state first-principles calculations. We find the calculations agree with experiments, and the effect of the $\gamma$-ray-induced excitation is significantly different from the effects of defect charge state and temperature. We propose a diffusion-based qualitative explanation of the mechanism of positive/negative ISE in NPN/PNP BJTs in the end.
\end{abstract}

\maketitle

\section{Introduction}
\label{sec:intro}
The ionizing damage (ID)\cite{NSREC2014} and the displacement damage (DD)\cite{NSREC2013} are the two major types of irradiation damages in semiconductor transistors. The ID results from ionizing irradiations such as the $\gamma$ ray, X ray, electron and ion irradiations, and the DD results from impinging energetic particles such as the neutron and ion irradiations. Both are characterized by the macroscopic electrical performances of transistors, such as the base current or the reciprocal of the current gain in bipolar junction transistors (BJT). The degradation of these macroscopic properties is due to irradiation-induced defects.

Only a few types of defects affect the macroscopic performance of transistors significantly. For Silicon-based transistors, the interface traps (IT) at the Si/SiO$_2$ interface and oxide traps (OT) in SiO$_2$\cite{NSREC2014} contribute the most to the ID, and the displacement defects such as divacancy V$_2$, vacancy-oxygen VO, divacancy-oxygen V$_2$O and defect clusters in the bulk Silicon contribute the most to the DD\cite{NSREC2013,W66,KP15,SZLZ19}. Thus the regions mainly affected by the ID and the DD do not overlap. Although the ionizing irradiation can affect bulk Silicon as well, these effects are less important than the IT and the OT, so they are usually not associated with the ID. Examples of such effects in bulk Silicon include the well-known low-temperature carrier-enhanced migration of Silicon self-interstitials Si$_\mathrm{i}$\cite{BJ84,BJ84b,HKJS99,JCGB09,ALSM21} and recombination-enhanced defect reactions\cite{LK74,WTK75,K76,K78}, and the generation of V$_2$, VO and V$_2$O\cite{BB82,YPPZ92,E93,SMAM04,MMAA05}.

One may expect the overall damage of ID and DD to be additive as they mainly affect different regions \revise{of the transistor}, but this is not true due to the irradiation synergistic effect (ISE)\cite{BSSB01,BSSF02,GLK04,LGLZ10,LLGR12,LLRG12,LLY15,WCYJ16,SZLZ19,SZCL20,SW20}. After simultaneous\cite{BSSB01,BSSF02,LGLZ10,LLGR12,LLRG12,LLY15} or sequential\cite{GLK04,LLGR12,LLRG12,WCYJ16,SZLZ19,SZCL20,SW20} displacement and ionizing irradiations, the overall damage \revise{to the transistor} can be either enhanced (positive ISE)\cite{LLRG12,LLY15,SZCL20,SW20} or reduced (negative ISE)\cite{BSSB01,BSSF02,GLK04,LLGR12,WCYJ16,SZLZ19,SW20} comparing with the sum of the individual ID and DD. The magnitude of the ISE varies a lot in different types of transistors. For Silicon-based BJTs, the type of the ISE appears to be correlated with the P/N-type\cite{GLK04,LLGR12,LLRG12,LLY15,WCYJ16,SZLZ19,SZCL20,SW20} of the base region Silicon, but there are also counter examples\cite{BSSB01,LGLZ10}. We only discuss the ISE of sequential neutron-$\gamma$ irradiations\cite{GLK04,LLGR12,LLRG12,SZLZ19,SZCL20,SW20} in Silicon-based BJTs in this paper.

The mechanism of the ISE is not well-understood. Traditionally it is explained by a space charge model\cite{BSSB01,BSSF02,SW20} where the charged OT in SiO$_2$ affects the Shockley-Read-Hall (SRH) recombination current in bulk Silicon through Coulomb interaction, but recent studies\cite{SZLZ19,SZCL20,SW20} pointed out its inability in explaining the dose-rate dependence of the ISE. The defect-based model of Ref. \onlinecite{SW20} solves this problem and achieves quantitative agreement with ISE experiments, where the ISE is explained by the enhanced migration and reactions of displacement defects in bulk Silicon due to $\gamma$-ray irradiation. \revise{However, the defect migration/reactions in the model are inferred from apparent functional forms of the data and lack validation.}

Due to restrictions on the temporal and spatial resolutions of the experimental characterization techniques, the most feasible approach to validate and go beyond the \revise{defect-based model} is computer simulations. The simulation of the ISE requires calculating the effect of $\gamma$-ray irradiation on neutron-generated displacement defects in bulk Silicon, but this cannot be directly represented in first-principles methods. Previous studies of the ID assume equivalence between the effect of the $\gamma$-ray irradiation and Fermi level changes\cite{DAKF14,KP15,ZZLL18,SW20}. This indirect approach is unsatisfactory for the following reasons. (1) The relation between the radiation dose and the Fermi level change is unknown, so quantitative analysis is impossible. (2) The charged system due to Fermi level change is different in nature from the neutral electronic excitations due to $\gamma$-ray irradiations. (3) The Fermi level is an equilibrium property, but the local system under $\gamma$-ray irradiation is far from equilibrium. (4) The effect of doping is also modeled by Fermi level changes, whose physics is very different from that of $\gamma$-ray irradiation. Therefore it is more appropriate to model the effect of $\gamma$-ray irradiation with a different physical quantity.

In this paper, we model the effect of the $\gamma$-ray irradiation as electronic excitations. We propose a multiscale method for the simulation of the effect of $\gamma$-ray irradiation by combining the Monte Carlo simulation of high-energy processes\cite{GEANT4,GEANT4b} and density-functional theory (DFT)\cite{HK64,KS65,FNM03} calculations of $\gamma$-ray-induced excited states. We apply the method to verify the defect-based ISE model of Ref. \onlinecite{SW20} \revise{at the material level}, and show the necessity of treating electron excitations explicitly. We find that the $\gamma$-ray irradiation significantly alters the activation energies of migration and dissociation of various defects and their relative stabilities in a short time period after $\gamma$ photon incidence, which is different from the effect of temperature and defect charge state. The calculation results agrees with the aforementioned $\gamma$-ray-induced defect generation and recombination-enhanced defect reactions in bulk Silicon, and can better explain the low-temperature carrier-enhanced migration of Si$_\mathrm{i}$. We provide a diffusion-based explanation for the different types of the ISE based on calculation results.

\section{Brief summary of the ISE and the defect-based model}
\label{sec:summaryISE}
We first provide a brief summary of the ISE experiments and the defect-based model of Refs. \onlinecite{SZLZ19,SZCL20,SW20}, as the calculations in this work are based on and compared to these experiments and the model. In Refs. \onlinecite{SZLZ19,SZCL20,SW20}, both NPN (3DK9D transistor) and PNP (LM324N chip with 4 PNP transistors) samples are irradiated sequentially first by neutron and then by the $\gamma$ ray. Experiments are carried out with all combinations of 4 neutron fluences ($F=0\sim 6\times10^{12}$ cm$^{-2}$ for NPN, $F=0\sim 5\times10^{13}$ cm$^{-2}$ for PNP) and 2 $\gamma$-ray dose rates [$R=10$ mrad(Si)/s and 10 rad(Si)/s for NPN, $R=2.2$ mrad(Si)/s and 10 rad(Si)/s for PNP], and data is recorded in each experiment at various $\gamma$-ray doses [$D=0\sim 5$ krad(Si)]. One of the major effect of irradiation damages is on the lifetime of minority carriers due to the generation of recombination centers\cite{NSREC2013,NSREC2014}, so the irradiation damage of a single BJT is usually characterized by the minority-carrier-dominated base current $\IB$. Since the LM324N chip contains multiple transistors, the damage is characterized by the input bias current $I_\mathrm{iB}$ which is proportional to $\IB$ of the input PNP transistors. We denote both $\IB$ and $I_\mathrm{iB}$ as $\IB$ in the following to simplify the notation.

$\IB$ has a strong sublinear dependence on the neutron fluence in neutron-only experiments, and an almost linear dependence on the $\gamma$-ray dose in $\gamma$-ray-only experiments. The summed damage is then $\IB(F)+k(R) D$, where $\IB(F)$ is the neutron-only $\IB$ at fluence $F$, $k(R)$ is the fitted slope of $\gamma$-ray-only $\IB$ at dose rate $R$, and $D$ is the $\gamma$-ray dose. $\IB$ of sequential neutron and $\gamma$-ray irradiation in NPN/PNP is higher/lower than this sum, which is denoted as positive/negative ISE. Experimental data suggests that $\IB(D)$ contains an exponentially decaying part and a linear part, and the difference between NPN and PNP is in the sign of the exponential part.

Based on various experimental results\cite{GLK04,LLGR12,LLRG12,LLY15,WCYJ16,SZLZ19,SZCL20,SW20}, Refs. \onlinecite{SZLZ19,SZCL20,SW20} associate the positive/negative type of the ISE with the P/N-type of the base region bulk Silicon. The strong dependence of the ISE on the neutron fluence suggests a defect-based mechanism where the concentrations of displacement defects change due to $\gamma$-ray-induced reactions. The electrically active defects for the ISE are traps for the majority carrier\cite{SW20}, so the active defect must have donor/acceptor levels in P/N-type Silicon.

$\IB$ is determined by the SRH recombination current, which is proportional to the concentrations of the electrically active defects. Minimal types of defects are considered in Ref. \onlinecite{SW20}, with V$_2$ and V$_2$O (0/+ and +/+2) as electrically active defects for P-type Silicon, and V$_2$ (0/- and -/-2), V$_2$O (0/- and -/-2) and VO for N-type Silicon. O$_\mathrm{i}$ is also included, which is a common impurity introduced by the growth process of Silicon\cite{L94}. Reactions involving these defects are assumed to be greatly enhanced by $\gamma$-ray irradiation, allowing $\partial/\partial t$ in rate equations to be replaced by $\partial/\partial D$. This assumption establishes a connection between defect concentrations and the irradiation dose.

Minimal types of reactions are included in the model, which are $\mathrm{V}_2+\mathrm{O}_\mathrm{i}\to\mathrm{V}_2\mathrm{O}$ and $\mathrm{V}_2+2\mathrm{Si}_\mathrm{i}\to0$ for P-type Silicon, and $\mathrm{V}_2+\mathrm{O}_\mathrm{i}\to\mathrm{V}_2\mathrm{O}$, $\mathrm{VO}+\mathrm{O}_\mathrm{i}\to\mathrm{VO}_2$ and $\mathrm{VO}+\mathrm{Si}_\mathrm{i}\to\mathrm{O}_i$ for N-type Silicon. $\IB$ is assumed to be approximately proportional to the concentrations of electrically active defects for simplicity. By solving the coupled reaction rate equations with proper boundary conditions, one arrives at the formula of $\IB$ in terms of $F$, $D$ and $R$. Remaining parameters are determined by fitting the ISE data.

The model has a good agreement with experiments and includes the dose rate effect, proving the validity of the defect-based mechanism. However, it also has the following flaws. (1) \revise{The defects and reactions are inferred from the apparent functional forms of $\IB$ and lacks verification.} (2) Some assumptions used in solving the rate equations might deivate from the practical situation. An example in the P-type model is that the concentrations of V$_2$ is assumed to be constant and proportional to $F$ for the annihilation of V$_2$ and Si$_\mathrm{i}$, but it is variable for the $\mathrm{V}_2+\mathrm{O}_i\to\mathrm{V}_2\mathrm{O}$ reaction. (3) Some of the defect levels in the model are controversial such as V$_2$O(+/+2)\cite{GRVM12}. (4) \revise{The model remains partly phenomenological even though the defects are specified, since the inferred attribution of defects lacks direct experimental verification, and the fitted model would still work even if the actual mechanism differs from the model as long as the final functional form has an exponentially decaying term plus a linear term.}

\section{Method}
\label{sec:method}
The $\gamma$ ray interacts with matter first through high-energy processes such as the Compton scattering and photoelectric effect, which results in secondary $\gamma$-ray or X-ray photons and a cascade of highly energetic electrons corresponding to excitations of core electrons.\cite{NMR12} Then the high-energy electrons de-excite through fluorescence, the Auger process and inelastic scattering, and eventually become what is commonly regarded as carriers, i.e., corresponding to excitations between valence and conduction bands. In the end, the extra carriers would vanish and return to the thermal equilibrium. This process cannot be simulated in entirety due to the large difference in the spatial and temporal scales of high- and low-energy processes. Technical limitations also hamper the simulation, such as the implicit treatment of core electrons in first-principles calculation for solids.

We simulate the interaction of the $\gamma$-ray with bulk Silicon in a multiscale manner. We first use the Monte-Carlo method\cite{GEANT4,GEANT4b} to simulate the high-energy processes until the termination of electron tracks, which yields the spatial distribution of excited electrons. We then carry out excited-state DFT calculations using the $\Delta$ self-consistent field ($\Delta$SCF)\cite{SW71,ZRB77} method and the Ehrenfest dynamics\cite{DTK72} with time-dependent density-functional theory (TDDFT)\cite{RG84,U12} to obtain properties of the displacement defects in the $\gamma$-ray-induced excited state.

\subsection{Monte-Carlo simulation of high-energy processes of the $\gamma$-ray irradiation}
\label{sec:method:geant4}
We use the Geant4 toolkit\cite{GEANT4,GEANT4b} for Monte-Carlo simulations. This simulation is for the high-energy processes of the $\gamma$-ray irradiation, but it need to incorporate the low-end of high-energy physics to be relevant to the irradiation damages of transistors. We therefore use the QBBC physics list\cite{IIMI12} together with the ``Livermore'' set of electromagnetic models as they are recommended for energy range below 1 GeV\cite{GEANT4App}. The atom de-excitation models and MicroElec low-energy electromagnetic models\cite{GICR21} of Geant4 are included in the simulations to describe the de-excitation process of the core excitations, and all the electrons are tracked without energy cutoff, so that the electrons at the end of their tracks can be interpreted as excited electrons on the conduction band. We treat the last-step electron energies and positions as energies above the conduction band minimum (CBM) and positions of the $\gamma$-ray-induced carriers.

\begin{figure}
\includegraphics[width=\columnwidth]{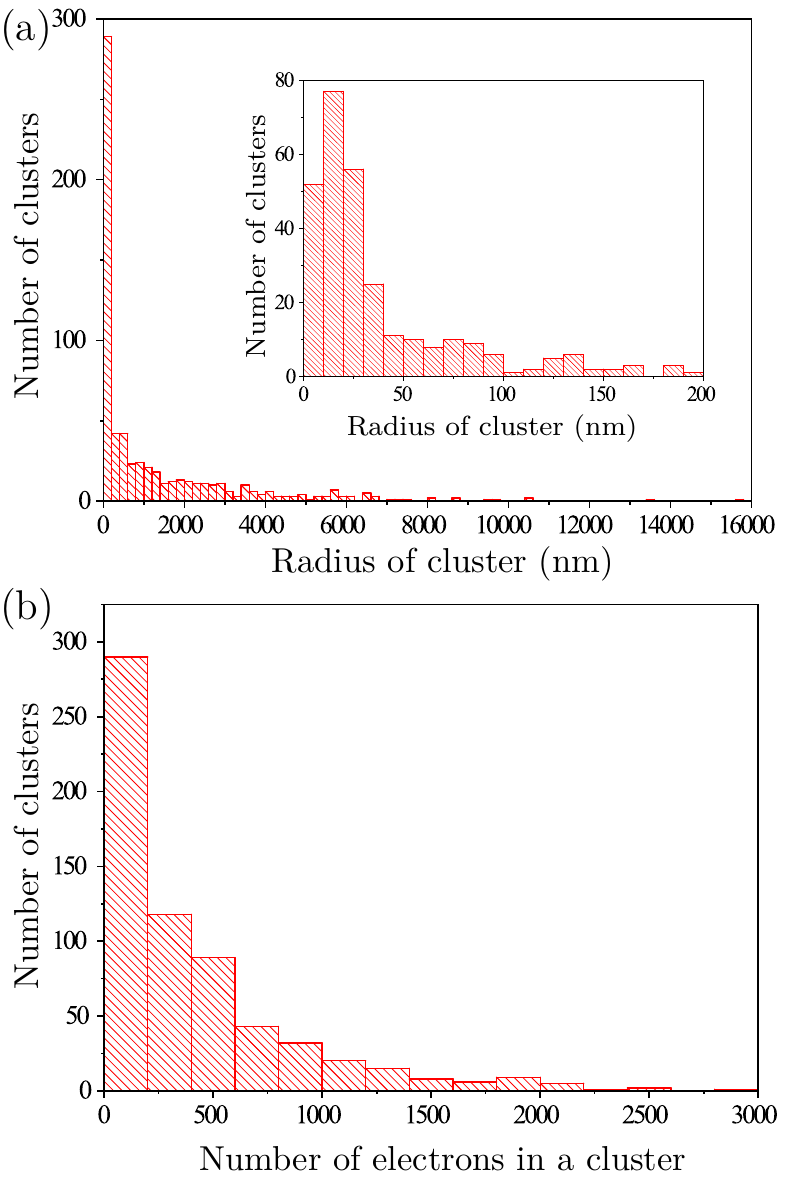}
\caption{(a) Radius and (b) size of electron clusters in Geant4 simulations. A total of 10 simulations are carried out. We use the agglomerative hierarchical clustering algorithm\cite{ALGLIB} to identify the clusters.}
\label{fig:ecluster}
\end{figure}

The setup of the simulations is described in the supplemental material\cite{supplemental}. The effect of the incidence of one $\gamma$ photon is highly local due to its high energy. This locality is verified by Geant4 simulations. We carry out 10 simulations to determine the average effect of one $\gamma$ photon on electronic excitations in bulk Silicon. Fig. \ref{fig:ecluster} shows that the last-step positions of the electrons aggregate in small clusters, and these clusters can be considered local electronic excitations. The locality can be seen from the very low average density of excited electrons as well, which is $5.5\times 10^{-15} $\AA$^{-3}$ for every 1 mrad $\gamma$-ray dose.

This locality allows us to use first-principles methods together with Geant4. We assume that the locally excited regions are uniformly distributed in the system at any given time, and we are interested in cases where the excitation happens in the vicinities of displacement defects. The sizes of the clusters are too big to be used directly in first-principles calculations, so we obtain the number of excited electrons by counting the max number of electrons in each cluster that can fit into the $3\times 3\times 3$ supercell of DFT (see supplemental material\cite{supplemental} for details). We would like to evaluate the largest impact of the $\gamma$-ray-induced excitation, so we average of the max number in each of the 10 simulations, and use 26 excited electrons in DFT calculations.

\subsection{First-principles calculations of excited states}
\label{sec:method:dft}
We use DFT to calculate the excited-state properties of displacement defects. All calculations are carried out on a $3\times 3\times 3$ supercell of bulk Silicon with the PBE generalized gradient approximation (GGA) exchange-correlation functional\cite{PBE96} and are spin-unpolarized and $\Gamma$-point-only. We use the Vienna \textit{ab inito} simulation package (VASP)\cite{KF96} for DFT and the PWmat software\cite{JFCW13,JCWF13} for TDDFT.

\begin{figure}
\includegraphics[width=\columnwidth]{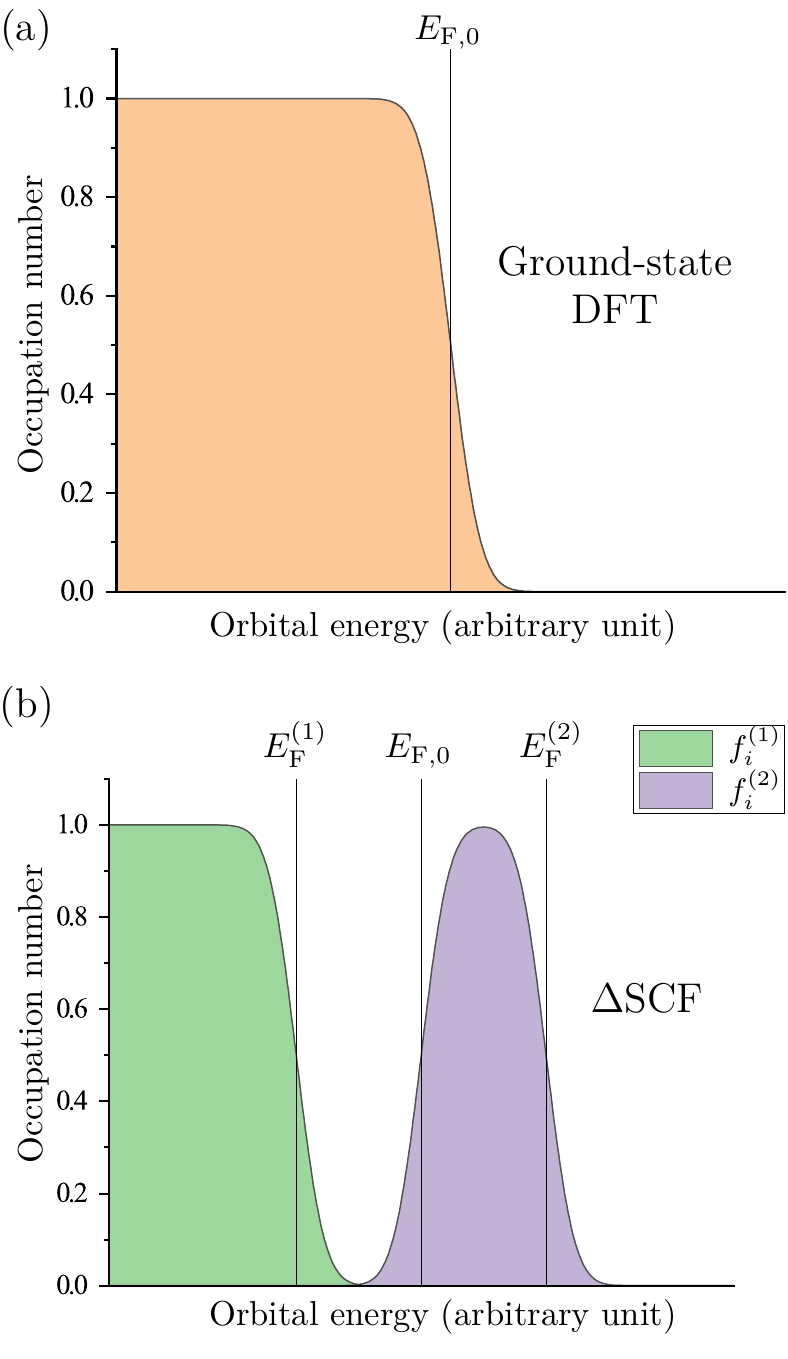}
\caption{Illustration of the ground-state and $\Delta$SCF occupation numbers.}
\label{fig:dscfoccnum}
\end{figure}

The energetics of typical displacement defects under $\gamma$-ray irradiation is calculated by the $\Delta$SCF method, whose low computational cost is essential for supercell calculations. $\Delta$SCF represents the excited state as a single Slater determinant with some occupied Kohn-Sham (KS) orbitals replaced by virtual ones, which is variationally optimized similar to the ground-state KS-DFT\cite{KS65}. The $\Delta$SCF occupation number $f_i$ of orbital $i$ is calculated by
\ben
f_i=f_i^{(1)}+f_i^{(2)},
\label{eqn:dscfocc}
\een
\begin{align}
\label{eqn:dscfocc1}
f_i^{(1)}&=\frac{1}{2}\left[1+\erf\left(\frac{E_{\mathrm{F},N_\text{exc}}^{(1)}-\epsilon_i}{\sigma}\right)\right],\\
\label{eqn:dscfocc2}
f_i^{(2)}&=\frac{1}{4}\left[1+\erf\left(\frac{\epsilon_i-E_{\mathrm{F},0}}{\sigma}\right)\right]\left[1+\erf\left(\frac{E_{\mathrm{F},N_\text{exc}}^{(2)}-\epsilon_i}{\sigma}\right)\right],
\end{align}
where $\epsilon_i$ is the orbital energy, $\sigma$ is the smearing factor, $E_{\mathrm{F},0}$ is the ground-state Fermi level of $N$ electrons, $N_\text{exc}$ is the number of excited electrons, and $E_{\mathrm{F},N_\text{exc}}^{(1)}$ and $E_{\mathrm{F},N_\text{exc}}^{(2)}$ are defined by $\sum_i f_i^{(1)}=N-N_\text{exc}$ and $\sum_i f_i^{(2)}=N_\text{exc}$. We use Gaussian smearing in Eqs. \ref{eqn:dscfocc1} and \ref{eqn:dscfocc2} to facilitate convergence. Fig. \ref{fig:dscfoccnum} compares $f_i$ of  Eq. \ref{eqn:dscfocc} with the ground-state ones.

Eq. \parref{eqn:dscfocc} only defines one specific excited state with $N_\text{exc}$ excited electrons. Although Geant4 simulations can yield the energy spectrum of electrons, we do not set $f_i$ accordingly due to it being incompatible with the DFT band structures and accuracy problems. See supplemental material\cite{supplemental} for details. We choose to use Eq. \parref{eqn:dscfocc} instead for its agreement with the classical picture of carriers in semiconductor physics, where excited electrons/holes filling up orbitals starting from the CBM/VBM respectively.

Unlike most $\Delta$SCF applications, Eq. \parref{eqn:dscfocc} does not define a low-lying excited state with one excited electron, leading to more profound convergence problem\cite{ZKSP16}. We use a large smearing parameter ($\sigma=0.3$ eV) to ensure convergence. The error due to smearing is small in many cases, but it sometimes lead to spurious geometries (see supplemental material\cite{supplemental}). We will check the effect of more sophisticated $\Delta$SCF convergence methods\cite{GBG08,PKZ15,HH20} in the future.

The dynamical properties of typical displacement defects under $\gamma$-ray irradiation are calculated by TDDFT Ehrenfest dynamics. The large smearing of $\Delta$SCF is unnecessary and not used in TDDFT. We use the highly efficient real-time TDDFT algorithm of Wang, et al\cite{WLW15} which allows a large time step of 0.1 fs, so the simulation time (ps level) can be relevant to ionic movements.

The initial state has ground-state KS orbitals and occupation numbers defined by Eq. \parref{eqn:dscfocc}. The initial ionic velocities are randomized according to the temperature. We use the canonical ensemble (NVT) in simulations. We also employ the Boltzmann factor method\cite{W20} which fixes the artificial heating of Ehrenfest dynamics and introduces decoherence. The energy loss due to the Boltzmann factor is compensated by the kinetic energy of nuclei in the direction of the corresponding excitation\cite{PWmatManual}. This allows us to incorporate the effect of the temperature, non-radiative de-excitation and electron-phonon coupling in simulations.

\section{Results and discussion}
\label{sec:results}
We verify the defect-based model of the ISE of Ref. \onlinecite{SW20} in the following. Our calculations confirm the assumption of Ref. \onlinecite{SW20} that the $\gamma$-ray irradiation significantly enhance the reactions/migrations of electrically active defects. The dose rate effect on the reaction rate constant can be explained as well. We find the $\gamma$-ray-induced electronic excitation only exist in a short time period, which justifies the assumed equivalence between the $\gamma$-ray dose and the reaction time in Ref. \onlinecite{SW20}. Unlike Ref. \onlinecite{SW20}, however, we also find that more types of defects are involved in the ISE, and not all defect reactions/migrations are enhanced under $\gamma$-ray irradiation. By estimating whether a reaction is enhanced/weakened according to the migration activation energies of the reactants, we provide a simple explanation of the positive/negative ISE in P/N-type Silicon based on the status of reactions whose reactants and products have different electrical activeness. Aside from verifying Ref. \onlinecite{SW20}, our results also explains the experimentally observed defect generation in Silicon after ionizing irradiation, verifies the theory of the recombination-enhanced defect reactions, and can explain the carrier-enhanced migration of Si$_\mathrm{i}$ better than existing charge-based theory.

\subsection{V$_2$ migration and dissociation}
\label{sec:results:exccalc:V2}
V$_2$ is one of the main type of displacement defects\cite{NSREC2013}. We first perform calculations on V$_2$ to have an overall idea of the behavior of displacement defects in $\gamma$-ray-induced excited states. V$_2$ is stable and largely immobile at room temperature and the ground state due to high activation energies for migration and dissociation (denoted as $E_{\mathrm{a},\mathrm{V}_2}^\text{mig}$ and $E_{\mathrm{a},\mathrm{V}_2}^\text{dis}$), which are 1.3 eV (by experiment\cite{MMAA05} and computation\cite{HG02}) and 2.0 eV (by computation\cite{HG02}) respectively. This stability agrees with our Born-Oppenheimer (BO) molecular dynamics (MD) NVT simulation of V$_2$ with ground-state DFT at 300 K, where all atoms vibrate around equilibrium positions during the 1 ps simulation, and V$_2$ does not move or dissociate according to the Wigner-Seitz defect analysis\cite{S10}.

In comparison, we carry out an Ehrenfest dynamics NVT simulation with ionic positions and velocities exactly the same as the BOMD for 350 fs. 26 electrons are initially excited according to Sec. \ref{sec:method:geant4}. The behavior of V$_2$ under the $\gamma$-ray-induced excited state is drastically different from that of the ground state. At 99.5 fs, V$_2$ spontaneously transforms to V-Si-V, which is the transition state for V$_2$ migration\cite{HG02}. At 103.3 fs, the defect structure transforms to V-Si-Si-V, which is the first dissociation step of V$_2$\cite{HG02}. At 193.4 fs, the defect becomes V-Si-V again. At 346.4 fs, the defect returns to the V$_2$ structure. Although the exact times does not mean much due to their dependence on the initial configuration, the time scale of these structural changes clearly indicate that $E_{\mathrm{a},\mathrm{V}_2}^\text{mig}$ and $E_{\mathrm{a},\mathrm{V}_2}^\text{dis}$ are much lower under $\gamma$-ray irradiation. This is verified by the following activation energy calculations.

\begin{figure}
\includegraphics[width=\columnwidth]{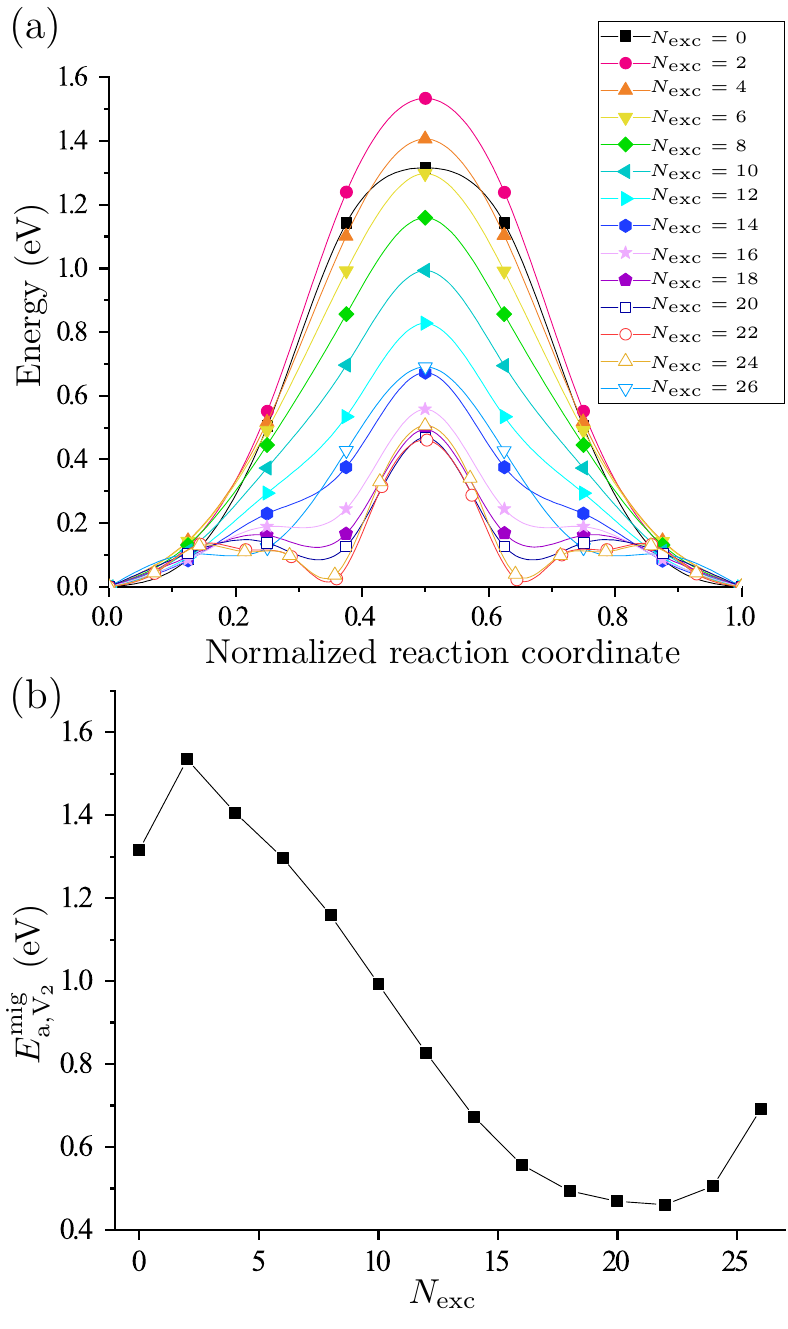}
\caption{(a) The total energy change along the V$_2$ migration path with different $N_\text{exc}$. The normalized reaction coordinate of the horizontal axis is the distance to the initial structure divided by the length of the path. (b) Plot of $E_{\mathrm{a},\mathrm{V}_2}^\text{mig}$ versus the $N_\text{exc}$.}
\label{fig:V2NEB}
\end{figure}

We calculate $E_{\mathrm{a},\mathrm{V}_2}^\text{mig}$ under $\gamma$-ray irradiation with the nudged elastic band (NEB)\cite{JMJ98} method. Fig. \ref{fig:V2NEB}(a) plots the energy along the migration path. For the ground state, we obtain $E_{\mathrm{a},\mathrm{V}_2}^\text{mig}=1.3$ eV as in Refs. \onlinecite{MMAA05,HG02}. With 26 excited electrons,  $E_{\mathrm{a},\mathrm{V}_2}^\text{mig}$ lowers to 0.7 eV. Fig. \ref{fig:V2NEB}(b) shows that $E_{\mathrm{a},\mathrm{V}_2}^\text{mig}$ has a nonmonotonic relationship with $N_\text{exc}$, and can be even higher than that of the ground state. Nevertheless, V$_2$ migration is enhanced in most cases under $\gamma$-ray irradiation.

V$_2$ dissociation is enhanced under $\gamma$-ray irradiation as well. We obtain $E_{\mathrm{a},\mathrm{V}_2}^\text{dis}=1.9$ eV for the ground state similar to Ref. \onlinecite{HG02}, and $E_{\mathrm{a},\mathrm{V}_2}^\text{dis}=0.8$ eV for 26 excited electrons. Due to the $\Delta$SCF convergence problem (see supplemental material\cite{supplemental} for details), we only discuss migration in the following.

The diffusivity $D$ of V$_2$ is related to $E_\mathrm{a}^\text{mig}$ by
\ben
D=D_0 \exp[-E_\mathrm{a}^\text{mig}/(k_\mathrm{B} T)],
\label{eqn:diffusivity}
\een
where $k_\mathrm{B}$ is the Boltzmann constant, $T$ is the temperature, and $D_0$ is the prefactor containing contributions from the entropy and the attempt frequency\cite{BHC98}. Fig. \ref{fig:V2NEB}(b) indicates that the ground-state and excited-state diffusivities can differ by many orders of magnitude. The same difference in $D$ can be achieved in the ground-state by increasing the temperature. Considering de-excitation, we estimate that the excited state of 26 electrons is approximately equivalent to a twofold to threefold increase of temperature shortly after the $\gamma$ photon incidence. This equivalence is confirmed in a ground-state BOMD simulation with elevated temperature: At 900 K, V$_2$ transforms to the transition state of migration (V-Si-V) at 88.9 fs, which is quite close to the 99.5 fs of the Ehrenfest dynamics simulation. We neglect $D_0$ in this estimation since its effect is much smaller: the entropy contribution does not change much as the migration mechanism stays the same, and the attempt frequency contribution calculated by the Vineyard method\cite{V57,KL06} only changes by one order of magnitude (see Supplemental material\cite{supplemental} for details).

The above equivalence of temperature and excitation is for the diffusivity alone, and the excited-state dynamics is quite different from that of simple heating, even though the $\gamma$-ray irradiation does have temperature effects\cite{NSREC2014}. We calculate the temperature of the system using atomic kinetic energies, which in the entire excited-state Ehrenfest dynamics simulation is in the $200\sim 450$ K range, so there is no obvious global heating of the system. Local heating can be excluded as well by checking the ratio between the maximum and averaged atomic kinetic energy, whose range is quite similar for the ground state ($3\sim 9.2$ in 300 K BOMD) and for excited state ($2.3\sim 8.4$ with 26 excited electrons). The $\gamma$-ray-induced excitation and temperature changes have different effects, and there is no equivalence between them in general. 

\begin{figure}
\includegraphics[width=\columnwidth]{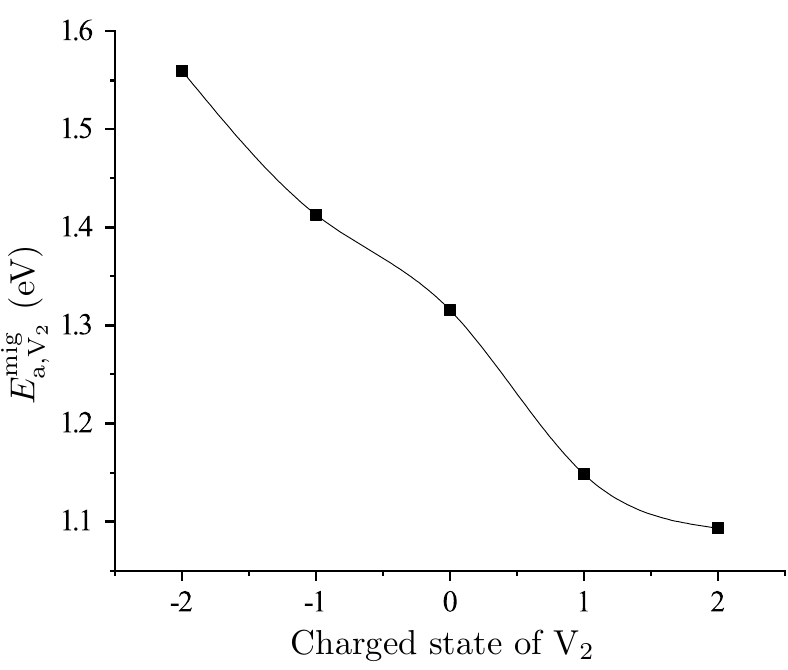}
\caption{$E_{\mathrm{a},\mathrm{V}_2}^\text{mig}$ plotted against the charged state of V$_2$.}
\label{fig:V2NEBCharged}
\end{figure}

The effect of $\gamma$-ray irradiation is commonly represented as Fermi-level changes. This can be checked by the Bader charge analysis\cite{HAJ06} of the Ehrenfest dynamics simulation. The Bader charges of Si atoms around V$_2$ are in the range of $2.79\sim 4.14$ and most of the times smaller than 4 (the number of valence electrons of the Si pseudopotential), suggesting the charge state of V$_2$ being $0\sim +2$. This is reasonable since the charge state of V$_2$ can be $-2\sim +2$ in literature.\cite{KP15,HEF97} Fig. \ref{fig:V2NEBCharged} shows $E_{\mathrm{a},\mathrm{V}_2}^\text{mig}$ of all charge states calculated by changing the Fermi level, and $E_{\mathrm{a},\mathrm{V}_2}^\text{mig}$ decrease monotonically with the charge state. This may appear to explain the enhanced V$_2$ migration, but the magnitude of the change is too small to explain the time scale in Ehrenfest dynamics. Therefore simple Fermi-level changes does not fully capture the effects of $\gamma$-ray irradiation, and $\gamma$-ray-induced excitations must be treated explicitly.

\begin{figure}
\includegraphics[width=\columnwidth]{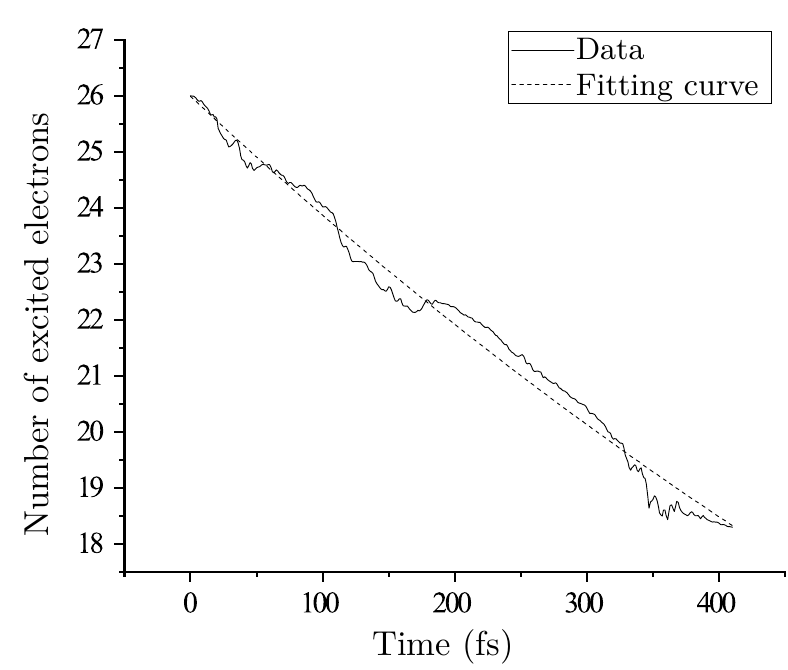}
\caption{$N_\text{exc}$ of the Ehrenfest dynamics simulation of V$_2$ plotted against time. See supplemental material\cite{supplemental} for details of the calculation of $N_\text{exc}$.}
\label{fig:TDNumExcit}
\end{figure}

We estimate the characteristic time of de-excitation using the time-dependent $N_\text{exc}$ in Fig. \ref{fig:TDNumExcit}. By fitting Fig. \ref{fig:TDNumExcit} with an exponentially decaying function, we obtain a 0.8 ps half-life for the $\gamma$-ray-induced excitation. The high-energy part of the de-excitation of Geant4 simulations takes 135 ps in average. Comparing with the lifetime of thermally excited minority carriers ($10^5\sim 10^7$ ps in Silicon at 300 K\cite{G69}), the de-excitation process of the $\gamma$-ray-induced excitation is much shorter. Therefore defect migration and reactions are only enhanced in a short duration after the incidence of a $\gamma$ photon. This justifies replacing $\partial/\partial t$ by $\partial/\partial D$ in the defect-based model of the ISE as shown in Sec. \ref{sec:summaryISE}. This is probably connected with the ISE dose rate effect as well, since the local system in the vicinity of a defect can remain excited with a sufficiently high $\gamma$-ray dose rate, and the average $N_\text{exc}$  of the local system would depend on the dose rate. The ISE dose rate effect is beyond the scope of this paper and will be studied in a future work.

Our Ehrenfest dynamics simulation verifies the theory of recombination-enhanced defect reactions under ionizing irradiation\cite{K78}. In the theory, the defect migration, dissociation and reactions forbidden at the temperature of the crystal become allowed by localized phonon excitations at the defect site due to electron-hole recombination. This has a direct correspondence to our simulation, where the recombination corresponds to the decreasing $N_\text{exc}$ over time, and the localized phonon allowing forbidden processes corresponds to V$_2$ structural changes without increasing the overall temperature. This theory can be seen as an alternative description of the non-radiative de-excitation process, and is incorporated in our Ehrenfest dynamics simulation as the compensation to the Boltzmann correction energy loss.

\begin{figure}
\includegraphics[width=\columnwidth]{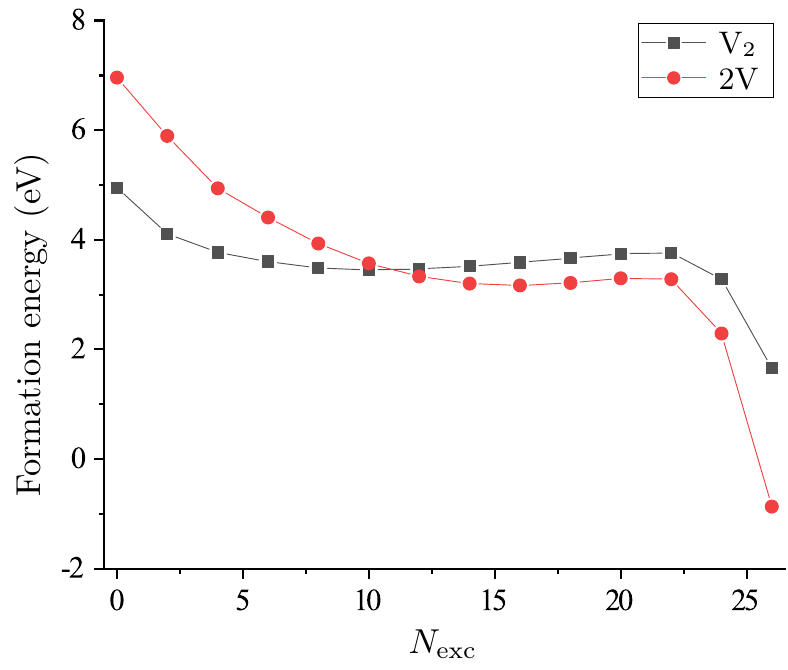}
\caption{Formation energies of V$_2$ and 2V plotted against $N_\text{exc}$.}
\label{fig:V2EF}
\end{figure}

At the ground state, V$_2$ is more stable than the monovacancy, and the mechanisms of V$_2$ migration and dissociation are different. The transition state for migration is V-Si-V\cite{HG02}, while V$_2$ dissociates to V-Si-Si-V with a one-step mechanism\cite{HG02} without going through V-Si-V. A different behavior in $\gamma$-ray-induced excited state is seen in the Ehrenfest dynamics simulation, where V-Si-V exists for a substantially long time for an unstable transition state, and can transform to and from V-Si-Si-V. The higher stability of V-Si-V in excited state can be checked by an Ehrenfest dynamics NVT simulation without the Boltzmann factor, where V-Si-V exists for a longer time (337 fs) because the de-excitation happens much slower in this simulation. This indicates that V$_2$ is less stable in the $\gamma$-ray-induced excited state than in the ground state, which can be verified by formation energies. Fig. \ref{fig:V2EF} plots the formation energies\cite{W04} of V$_2$ and two isolated vacancies (2V), and we find that 2V becomes more favorable than V$_2$ as $N_\text{exc}$ increase.

The formation energy of V becomes negative with 26 excited electrons, indicating that the solid becomes unstable at this point. This should not be a problem since the excitation is local and de-excitation is fast. Nevertheless, we can expect that more vacancies are generated under $\gamma$-ray irradiation, considering that the formation of the Si$_\mathrm{i}$-V pair only takes 410 fs\cite{TCZD97}. This agrees with experiments\cite{BB82,YPPZ92} where V$_2$ and VO are generated in bulk Silicon after ionizing irradiations.

The fitting parameters of the defect-based ISE model of Ref. \onlinecite{SW20} indicate that the reaction $\mathrm{V}_2+\mathrm{O}_\mathrm{i}\to\mathrm{V}_2\mathrm{O}$ happens faster at higher $\gamma$-ray dose rate (\revise{but not fast enough to be proportional to the dose rate change, and the dose rate effect is more complicated than it appears,} see supplemental material\cite{supplemental}). As the higher dose rate is equivalent to more excited electrons in average, this can be explained by the increase of V$_2$ diffusivity of V$_2$ due to the decrease of $E_{\mathrm{a},\mathrm{V}_2}^\text{mig}$.\cite{W58,WLZL19}. The change of the formation energy of $V$ and $V_2$ may also influence the fitted reaction rate parameters, as the increased concentration of V$_2$ makes the reaction more likely to happen.

\subsection{Migration of other defects}
\label{sec:results:exccalc:other}
Table \ref{table:migrationbarrier} lists $E_\mathrm{a}^\text{mig}$ of other displacement defects (V, VO, V$_2$O, Si$_\mathrm{i}$) and O$_i$ with 26 excited electrons. The calculation for VO is set up according to the six-member-ring mechanism of Ref. \onlinecite{FT05}, where $E_\mathrm{a}^\text{mig}$ and $E_\mathrm{a}^\text{dis}$ are related and very close to each other. The migration mechanism of V$_2$O is not reported in literature as far as we aware, so we assume that its migration and dissociation\cite{CJOB03} are closely related similar to the VO situation, since both V$_2$O and VO are composite defects.

\begin{table}
\caption{$E_\mathrm{a}^\text{mig,gs}$ and $E_\mathrm{a}^\text{mig,exc}$ with 26 excited electrons of various defects. $\rho$ denotes the estimated change in the diffusivity going from the ground state to the excited state.}
\begin{tabular}{cccc}
\hline\hline
 & $E_\mathrm{a}^\text{mig,gs}$ & $E_\mathrm{a}^\text{mig,exc}$ & \\
\raisebox{1.5ex}[0pt]{Defect} & (eV) & (eV) & \raisebox{1.5ex}[0pt]{$\displaystyle\rho=\exp\left[-\frac{E_\mathrm{a}^\text{mig,exc}-E_\mathrm{a}^\text{mig,gs}}{k_\mathrm{B} T}\right]$}\\
\hline
V$_2$ & 1.3 & 0.54 & $3.0\times 10^{10}$\\
V & 0.070 & 0.24 & $1.2\times 10^{-3}$\\
VO & 1.8 & 0.31 & $1.8\times 10^{24}$\\
V$_2$O\footnotemark[1] & 1.6 & 1.1 & $2.9\times 10^8$\\
Si$_\mathrm{i}$ & 0.29\footnotemark[2] & 0.84\footnotemark[3] & N/A\footnotemark[4]\\
O$_\mathrm{i}$ & 2.1 & 0.37\footnotemark[5] & $4.2\times 10^{28}$\\
\hline\hline
\end{tabular}
\label{table:migrationbarrier}
\footnotetext[1]{Lists $E_\mathrm{a}^\text{dis}$, see explanation in the main text.}
\footnotetext[2]{The barrier between the split structure to the distorted hexagonal structure according to Ref. \onlinecite{JCGB09}.}
\footnotetext[3]{The barrier between the cross structures. See explanation in main text.}
\footnotetext[4]{$E_\mathrm{a,Si_i}^\text{mig}$ is irrelevant to diffusivity change due to carrier-enhanced migration. See explanation in main text.}
\footnotetext[5]{The barrier between the T-shaped structures. See explanation in main text.}
\end{table}

\begin{figure}
\includegraphics[width=\columnwidth]{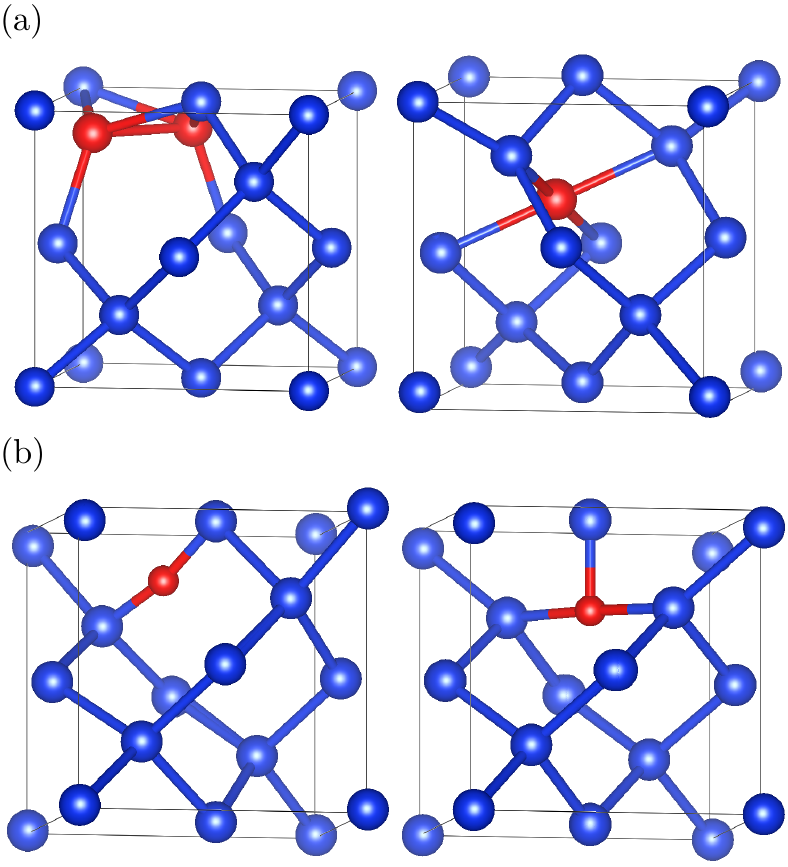}
\caption{The most stable structures of (a) Si$_\mathrm{i}$ and (b) O$_\mathrm{i}$ in the ground state and in the excited state with 26 excited electrons.}
\label{fig:structure}
\end{figure}

$E_\mathrm{a}^\text{mig}$ is different for the ground state and the excited state. Although the excitation due to one $\gamma$ photon only lasts for a short time, it is enough for the change in $E_\mathrm{a}^\text{mig}$ to be visible, considering the agreement between the dynamics of V$_2$ and the lowered activation energies in Sec. \ref{sec:results:exccalc:V2}. However, the situation is complicated by structural changes of some defects in the excited state. Table \ref{table:SiiOiEnergies} shows that the most stable ground-state and excited-state structures (Fig. \ref{fig:structure}) of Si$_\mathrm{i}$ and O$_\mathrm{i}$ are different, where the excited-state ones are discovered in NEB calculations. These excited-state structures are not reported in literature, which is expected since these structures are either unstable or energetically unfavorable in the ground state, and the $\gamma$-ray-induced excitation decays very fast. We show in the following that the structural difference explains the low-temperature carrier-enhanced migration of Si$_\mathrm{i}$.\cite{BJ84,BJ84b,HKJS99,JCGB09,ALSM21}

\begin{table}
\caption{The energies of different structures of Si$_\mathrm{i}$ and O$_\mathrm{i}$ in the ground state and the excited state with 26 excited electrons calculated with a $3\times 3\times 3$ supercell.}
\begin{tabular}{cccc}
\hline\hline
Defect & Structure & $E^\text{gs}$ (eV) & $E^\text{exc}$ (eV)\\
\hline
Si$_\mathrm{i}$ & Hexagonal site & -1170.7343 & -1139.8213\footnotemark[1]\\
Si$_\mathrm{i}$ & Tetrahedral site & -1170.7469 & -1139.7466\\
Si$_\mathrm{i}$ & Split & -1170.7761 & -1139.3214\\
Si$_\mathrm{i}$ & Cross & Unstable\footnotemark[2] & -1140.6234\\
O$_\mathrm{i}$ & Si-O-Si & -1176.2712 & -1142.7922\\
O$_\mathrm{i}$ & T-shape & -1174.1217 & -1143.6703\\
\hline\hline
\end{tabular}
\label{table:SiiOiEnergies}
\footnotetext[1]{Becomes the distorted hexagonal structure\cite{JCGB09}.}
\footnotetext[2]{Unstable for less than 10 excited electrons.}
\end{table}

The low-temperature (<65 K) migration of Si$_\mathrm{i}$ is known to be enhanced under ionizing irradiation\cite{BJ84,BJ84b,HKJS99,JCGB09,ALSM21}, where its diffusivity is much higher than extrapolated from high temperature data. Because of this, the annihilation of V and Si$_\mathrm{i}$ is enhanced at low temperature, explaining the low-temperature irradiation damage recovery\cite{JCGB09}. The enhanced migration was explained by the most stable Si$_\mathrm{i}$ structure being dependent on the charge state\cite{BJ84,BJ84b}, which is the split structure for Si$_\mathrm{i}(0)$ and the distorted hexagonal structure for Si$_\mathrm{i}(+1)$\cite{JCGB09}. The migration of Si$_\mathrm{i}$ is barrier-less\cite{JCGB09} as its charge state changes back and forth due to ionizing irradiation, resulting in a very small 0.065 eV $E_\mathrm{a,Si_\mathrm{i}}^\text{mig}$\cite{HKJS99} obtained from experiments. However, the charge state of Si$_i$ would need to oscillate for this to work, but the driving force of this oscillation is not explained. Although Ref. \onlinecite{JCGB09} proposes Si$_\mathrm{i}(+1)$ being a hole trap in P-type Silicon as the driving force, it cannot be applied to N-type Silicon which has carrier-enhanced migration of Si$_\mathrm{i}$ as well\cite{BJ84}. This charge-based theory also does not explain why the annihilation of V and Si$_i$ above 65 K is more difficult.\cite{JCGB09}

We reinvestigate the low-temperature carrier-enhanced migration of Si$_\mathrm{i}$ in the viewpoint of $\gamma$-ray-induced excitations. The most stable structures of the ground-state (split) and excited-state (cross) Si$_\mathrm{i}$ are different, and there is no barrier in the excited state going from the split structure to the cross structure. Similar to the charge-based theory, Si$_i$ can migrate to neighboring sites as the structure change spontaneously due to $\gamma$-ray-induced excitation and the following de-excitation. Unlike the charge-based theory, there is no problem with the driving force since the de-excitation happens naturally. This excitation-based explanation does not rely on the P/N-type of Silicon.

The difference between high- and low-temperature migration of Si$_\mathrm{i}$ under ionizing irradiation can be explained by the difference in the de-excitation time. Although the structural transformation of Si$_\mathrm{i}$ due to excitation and de-excitation is spontaneous, it still takes time and cannot happen instantly. The de-excitation half-lives at 300 K and 1500 K are respectively 4 ps and 0.1 ps calculated with Ehrenfest dynamics simulations of Si$_\mathrm{i}$. These values are less accurate as the simulations only run for 100 fs due to high computational cost, but the trend is enough to prove our point. As the excitation half-life decreases with the increasing temperature, it would eventually become less than the time needed for the spontaneous structure transformation, preventing the carrier-enhanced migration. The high-temperature diffusion is then determined by $E_\mathrm{a}^\text{mig,gs}$ if the $\gamma$-ray dose rate is not too high. This explains the weak temperature dependence of the carrier-enhanced migration of Si$_\mathrm{i}$\cite{HKJS99}, since it can happen as long as the de-excitation time is longer than the time necessary for structural transformation.

The most stable structure of O$_\mathrm{i}$ also changes with excitation according to calculation. Its migration should be enhanced as well following the same logic. Although Refs. \onlinecite{NTL83,OBNT84} report irradiation-enhanced diffusion of oxygen in Silicon, experimental evidences favor a mechanism involving VO instead\cite{OBNT84}, which agrees with the lowered $E_\mathrm{a,VO}^\text{mig,exc}$ in Table \ref{table:migrationbarrier}. The difference between Si$_\mathrm{i}$ and O$_\mathrm{i}$ is probably due to the cross structure of Si$_\mathrm{i}$ being unstable in the ground state while the T-shaped structure of O$_\mathrm{i}$ remains a local minimum, so the structure transformation of Si$_\mathrm{i}$ may happen faster than O$_\mathrm{i}$ during de-excitation.

\subsection{A diffusion-based explanation of the ISE}
\label{sec:results:ISEexplanation}
The rate equations in the defect-based model of the ISE\cite{SW20} have a similar form as those in liquid solutions. For example, the rate equation of $\mathrm{V}_2+\mathrm{O}_\mathrm{i}\to\mathrm{V}_2\mathrm{O}$ is written as $\partial c_{\mathrm{V}_2\mathrm{O}}/\partial t=k c_{\mathrm{V}_2}(t)c_{\mathrm{O}_\mathrm{i}}(t)$, where $c(t)$ is the concentration at time $t$ and $k$ is the rate constant. The concentrations of reactants are proportional to the collision probability of reactants and dominate the reaction rate in solutions. On the contrary, collisions happen scarcely in solids due diffusion being difficult, so the rate constant of a defect reaction should be dominated by diffusivities of reactants in the prefactor.\cite{W58,WLZL19} Under $\gamma$-ray irradiation, defect diffusion can be enhanced or weakened according to Table \ref{table:migrationbarrier}, suggesting a correlation between the ISE and the diffusivity changes due to $\gamma$-ray-induced excitation. In the following, we provide a diffusion-based explanation of the positive/negative ISE of the P/N-type Silicon.

$\rho$ of Table \ref{table:migrationbarrier} estimates the diffusivity changes due to $\gamma$-ray irradiation. We do not take the prefactor into account as in Sec. \ref{sec:results:exccalc:V2}. We assume that the carrier-enhanced migration of Si$_\mathrm{i}$ does not happen as the experiments of the defect-based model\cite{SZLZ19,SZCL20,SW20} do not specifically control the temperature, so the migration of Si$_\mathrm{i}$ is always determined by $E_\mathrm{a}^\text{mig,gs}$ and not enhanced/weakened under $\gamma$-ray irradiation. We then estimate whether a defect reaction is enhanced or weakened under $\gamma$-ray irradiation based on $\rho$.\cite{W58,WLZL19}

Based on this estimation, we propose that the ISE is mainly due to reactions of displacement defects whose reactants and products have different electrical activeness. For a reaction with active reactants and inactive products, it being enhanced/weakened under $\gamma$-ray irradiation means the damage characterized by $\IB$ is reduced/enhanced, and the reaction contributes to the negative/positive ISE. The contrary goes for reactions with inactive reactants and active products. Other types of reactions can contribute to the ISE due to energy level and capture cross section differences of the reactants and products, but the effect would be weaker. We also ignore reactions with more than two reactants, which are unlikely to happen in solids.

The electrically active defects for the ISE are donors in P-type Silicon and acceptors in N-type Silicon.\cite{SW20} Since $\IB$ is measured after irradiation, only the ground-state defect levels are relevant. For P-type Silicon, V$_2$\cite{GRVM12}, V$_2$O\cite{GRVM12}, Si$_\mathrm{i}$\cite{ALSM21} and V$_2$O$_2$\cite{CJOB03,GRVM12b} are active, and VO\cite{PKB02}, O$_\mathrm{i}$\cite{MK94} and VO$_2$\cite{SSL86} are inactive. For N-type Silicon, V$_2$\cite{GRVM12}, VO\cite{MAHS02,SCNK04}, V$_2$O\cite{MAHS02,SCNK04} and V$_2$O$_2$\cite{SRCN03,CJOB03} are active, and Si$_\mathrm{i}$\cite{NOYT03,JCGB09}, O$_\mathrm{i}$\cite{MK94} and VO$_2$\cite{SSL86,PLLH01} are inactive. The concentration of the monovacancy V should be very low due to the stability of V$_2$, so it is not included. Table \ref{table:reaction} lists the estimated behaviors of eligible defect reactions under $\gamma$-ray irradiation.

\begin{table}
\caption{Estimated behaviors of defect reactions of displacement defects under $\gamma$-ray irradiation. Reactions enhanced/weakened under $\gamma$-ray irradiation are denoted with $\uparrow$/$\downarrow$. The annihilation of V and Si$_\mathrm{i}$ is not included since the low-temperature carrier-enhanced migration of Si$_\mathrm{i}$ is assumed not happening.}
\begin{tabular}{cccc}
\hline\hline
& & & Influence\\
\raisebox{1.5ex}[0pt]{Reaction} & \raisebox{1.5ex}[0pt]{Type} & \raisebox{1.5ex}[0pt]{Status} & to the ISE\\
\hline
\multicolumn{4}{c}{1. P-type Silicon}\\
\hline
$\mathrm{V}_2\mathrm{O}+\mathrm{Si}_\mathrm{i}\to\mathrm{VO}$ & Active$\to$Inactive & $\downarrow$ & Positive\\
$\mathrm{V}_2\mathrm{O}_2+\mathrm{Si}_\mathrm{i}\to\mathrm{VO}_2$\footnotemark[1] & Active$\to$Inactive & - & N/A\\
$\mathrm{VO}+\mathrm{Si}_\mathrm{i}\to\mathrm{O}_\mathrm{i}$ & Active$\to$Inactive & $\uparrow$ & Negative\\
$\mathrm{VO}+\mathrm{VO}\to\mathrm{V}_2\mathrm{O}_2$ & Inactive$\to$Active & $\uparrow$ & Positive\\
\hline
\multicolumn{4}{c}{2. N-type Silicon}\\
\hline
$\mathrm{VO}+\mathrm{Si}_\mathrm{i}\to\mathrm{O}_\mathrm{i}$ & Active$\to$Inactive & $\uparrow$ & Negative\\
$\mathrm{VO}+\mathrm{O}_\mathrm{i}\to\mathrm{VO}_2$ & Active$\to$Inactive & $\uparrow$ & Negative\\
\hline\hline
\end{tabular}
\footnotetext[1]{The mechanism of V$_2$O$_2$ migration or dissociation is unavailable in literature as far as we aware, so it is assumed to be immobile due to the complex structure.}
\label{table:reaction}
\end{table}

The majority of the estimations in Table \ref{table:reaction} agrees with the experimentally observed positive/negative ISE of P/N-type Silicon. The type of the ISE is determined by defects with different electrical activeness in P- and N-type Silicon, which are VO and Si$_\mathrm{i}$ among the defects considered in this work.

This diffusion-based explanation is also applicable to the generation of V$_2$O after ionizing irradiation\cite{SMAM04,MMAA05}, since the reaction $\mathrm{V}_2+\mathrm{O}_\mathrm{i}\to\mathrm{V}_2\mathrm{O}$ is enhanced due to the diffusion of both V$_2$ and O$_\mathrm{i}$ being enhanced.

\section{Conclusion}
\label{sec:conclusion}
Electron excitation is closely related to both the ID and the DD in transistors. Explicitly treating excitations proved to be important in the DD\cite{LFHM16,LSDF20}, but previous first-principles studies of the ID employ an implicit qualitative approach assuming equivalence between the $\gamma$-ray irradiation and Fermi level changes. In this paper, we develop a multiscale method for the direct simulation of the effect of $\gamma$-ray irradiation in semiconductors by treating $\gamma$-ray-induced electronic excitations explicitly in first-principles calculations, with the excitation obtained from the Monte Carlo simulation of the high-energy interaction between the $\gamma$ ray and the material. We apply the method to the ISE in Silicon-based BJTs and verify the validity of the defect-based ISE model of Ref. \onlinecite{SW20}, showing that defect processes can be significantly altered under $\gamma$-ray irradiation. Our calculations agree with the experimentally observed defect generation in Silicon after ionizing irradiation, the recombination-enhanced defect reactions, and can better explain the low-temperature carrier-enhanced migration of Si$_\mathrm{i}$. A simple explanation of the positive/negative ISE in P/N-type Silicon is provided based on calculation.

The ISE is far from being well-understood and varies a lot in different transistors and devices. The experiments are heavily influenced by circuit-level and device-level factors that are difficult or impossible to be included in material-level first-principles calculations, which are necessary to go beyond \revise{existing phenomenological models and to verify the recent defect-based model\cite{SW20}}. Because of the complexity of the problem, this paper is only a preliminary step of understanding the mechanism of the ISE. The scope of this work is quite limited, since we only focus on the diffusion of a few types of defects and do not explicitly consider the effect of doping, $\gamma$-ray dose and dose rate. The underestimation of the formation energy by GGA also disallows direct comparison of diffusivity with experiments. Nevertheless, this work demonstrates the validity of the multiscale method, which allows direct computational study of other aspects of the ISE such as the effect of dopant atoms, the P/N-type of Silicon, the $\gamma$-ray dose and dose rate and so on. We will address these in future works.

\revise{
This work describes the non-equilibrium local transient effects around existing defects following the incident of one $\gamma$ photon. This microscopic picture complements the conventional quasi-Fermi energy model of semiconductors under ionizing irradiation. Considering the low defect concentrations, most irradiation-induced excitations would happen in regions far away from defects, generating extra carriers without structural changes in most cases. The generation, migration and recombination of these extra carriers eventually reach a quasi-equilibrium at a constant irradiation dose rate. This macroscopic quasi-equilibrium picture is described by the conventional quasi-Fermi energy model, which treats the irradiation effect in a mean-field manner. However, as we show in this paper, when a $\gamma$ photon deposit energy close to an existing defect, the microscopic non-equilibrium processes cannot be described by the averaged change in the quasi-Fermi levels, and they can have a significant impact on the mechanism of irradiation effects.
}

\begin{acknowledgments}
This work is supported by the National Natural Science Foundation of China under Grants Nos. 11804314 and 11804313.
\end{acknowledgments}

\section*{Conflict of Interest Statement}
The authors have no conflicts to disclose.

\section*{Data Availability Statement}
The data that support the findings of this study are available from the corresponding author upon reasonable request.

\section*{Author Contributions}
\textbf{Zeng-hui Yang:} Conceptualization (lead), Formal Analysis (lead), Methodology (lead), Software (lead), Writing/Original Draft Preparation (lead). \textbf{Yang Liu:} Validation (lead), Writing/Review \& Editing (equal). \textbf{Ning An:} Writing/Review \& Editing (equal). \textbf{Xingyu Chen:} Writing/Review \& Editing (equal).

\end{document}